\documentclass[aps,prl,twocolumn,superscriptaddress]{revtex4}
\usepackage{graphics,epsfig}

\newcommand{\beq}{\begin{equation}}
\newcommand{\eeq}{\end{equation}}
\newcommand{\bea}{\begin{eqnarray}}
\newcommand{\eea}{\end{eqnarray}}
\newcommand{\beas}{\begin{eqnarray*}}
\newcommand{\eeas}{\end{eqnarray*}}

\newcommand{\Jpar}{J_{\|}}
\newcommand{\Jperp}{J_{\bot}}

\newcommand{\pdag}{{\phantom{\dagger}}}

\newcommand{\nn}{\nonumber}

\begin{document}


\title{Crossover from Non--Equilibrium to Equilibrium Behavior in the Time--Dependent
Kondo Model}

\author{Dmitry Lobaskin}
\author{Stefan Kehrein} 
\affiliation{Theoretische Physik III -- Elektronische Korrelationen und
Magnetismus, Universit{\"a}t Augsburg, 86135 Augsburg, Germany} 


\date{\today}

\begin{abstract}
We investigate the equilibration of a Kondo model that is initially 
prepared in a non--equilibrium state towards its equilibrium 
behavior. Such initial non--equilibrium states
can e.g.\ be realized in quantum dot experiments with time--dependent
gate voltages. We evaluate the 
non--equilibrium spin--spin correlation
function at the Toulouse point of the Kondo model exactly and analyze
the crossover between non--equilibrium and
equilibrium behavior as the non--equilibrium initial state evolves
as a function of the waiting time for the first spin
measurement. Using the flow equation method we extend these
results to the experimentally relevant limit of small Kondo couplings.   
\end{abstract}


\maketitle



\textit{Introduction.--}
Equilibration in non--perturbative many--body problems
is not well--understood with many fundamental
questions still being unanswered. For example the crossover from 
a non--equilibrium initial state to equilibrium behavior after
a sufficiently long waiting time poses many interesting
questions that are both experimentally relevant and theoretically
of fundamental importance. From the experimental side there is
currrent interest in such questions related to 
transport experiments in quantum dots. Non--perturbative Kondo physics 
has been observed in quantum dots \cite{QuantumDots} and given
rise to a wealth of experimental and theoretical investigations.
Quantum dot experiments allow the possibility to systematically study the
effect of time--dependent parameters, like switching on the Kondo
coupling at a specific time and measuring the time--dependent
current. From the theoretical point of view this is related to
studying the time--dependent buildup of the non--perturbative
Kondo effect.

The combination of strong--coupling behavior and time--dependent
parameters makes such problems theoretically very
challenging.
Various methods like time--dependent NCA and
renormalized perturbation theory \cite{Nordlander}, 
the numerical renormalization group \cite{Costi},
bosonization and refermionization techniques \cite{Schiller},
etc.\  have allowed
insights and e.g.\ identified the time scale
$t_{\rm K}\propto \hbar/k_{\rm B} T_{\rm K}$ related to the Kondo
temperature~$T_{\rm K}$ with the relevant time scale for the
buildup of the Kondo resonance. Using the form factor approach
Lesage and Saleur could derive exact results for 
the spin expectation value $P(t)=\langle S_z(t) \rangle$
for a product initial state \cite{Lesage}.   
However, no exact results are
available regarding the crossover from non--equilibrium to
equilibrium behavior in this paradigm strong--coupling model
of condensed matter theory.

In this Letter we use bosonization and refermionization
techniques to calculate the zero temperature spin--spin correlation function
of the Kondo model at the Toulouse point \cite{Toulouse} exactly
for two non--equilibrium preparations: I)~The impurity spin is
frozen for time $t<0$. II)~The impurity spin is decoupled from
the Fermi sea for time $t<0$. We find a crossover between
non--equilibrium exponential decay and equilibrium algebraic decay
as one increases the waiting time for measuring the spin--spin
correlation function at time $t>0$. 
One concludes that zero temperature equilibration occurs exponentially
fast with a time scale set by the inverse Kondo temperature,
and a mixture of non--equilibrium and equilibrium behavior for
finite waiting times. Using the flow equation solution of
the Kondo model \cite{Hofstetter,Slezak} we then
extend these results away from the Toulouse point to the experimentally
relevant limit of small Kondo couplings in a systematic approximation.

\textit{Model.--}
The Kondo model describes the interaction of a
spin--1/2 degree of freedom $\vec S$ with a Fermi sea
\beq
H=\sum_{k,\alpha} \epsilon_k c^\dag_{k\alpha} c^\pdag_{k\alpha}
+\sum_i J_i \sum_{\alpha,\beta}
c^\dag_{0\alpha}\,  S^\pdag_i \, \sigma^{\alpha\beta}_i\:
c^\pdag_{0\beta} \ .
\label{Kondo_Hamiltonian}
\eeq
Here $c^\dag_{0\alpha}, c^\pdag_{0\alpha}$
is the localized electron orbital at the impurity site.
We allow for anisotropic couplings
$J_i=(\Jperp,\Jperp,\Jpar)$ and consider a linear dispersion
relation $\epsilon_k=v_F\,k$. Throughout this Letter
we are interested in the universal behavior in the scaling 
limit $\Jperp/v_F \rightarrow 0$.
The Kondo Hamiltonian can be mapped to the spin--boson model,
which is the paradigm model of dissipative quantum mechanics \cite{Leggett}. 
Our results
can be interpreted in both these model settings; in the sequel
we will focus on the Kondo model interpretation.

We study two non--equilibrium preparations:
I)~The impurity spin is frozen for time $t<0$ by a large
magnetic field term $h(t)S_z$ that is switched
off at $t=0$: $h(t)\gg T_{\rm K}$ for $t<0$ and $h(t)=0$ for $t\geq 0$.
II)~The impurity spin is decoupled from the bath degrees of freedom for
time $t<0$ (like in
situation~I we assume $\langle S_z(t\leq 0)\rangle=+1/2$)
and then the coupling is switched on at $t=0$: $J_i(t)=0$ for
$t<0$ and $J_i(t)=J_i>0$ time--independent for $t\geq 0$.
This situation is of particular interest in future quantum dot experiments
where the quantum dot is suddenly
switched into the Kondo regime by applying a time--dependent
voltage on a nearby gate \cite{Nordlander}.

The difference between these two 
initial states is that in~I) the electrons are
in equilibrium with respect to
the potential scattering induced by the frozen spin for $t<0$.
On the other hand in~II) the initial state of the electrons
is an unperturbed Fermi sea. We will later
see that both initial
states lead to the same spin dynamics.

A suitable quantity for studying equilibration 
is the symmetrized zero temperature
spin--spin correlation function
$C_{\rm I,II}(t_w,t)\stackrel{\rm def}{=} \frac{1}{2}
\langle \{S_z(t_w),S_z(t_w+t)\} \rangle_{\rm I,II}$.
In equilibrium this correlation function exhibits its well--known
$t^{-2}$-algebraic long time decay \cite{Leggett} and is, of course,
independent from the initial ({\em waiting}) time~$t_w$:
$C_{\rm eq}(t)=C_{\rm eq}(t_w,t)\propto t^{-2}$.

Exact results for the non--equilibrium spin dynamics have so far only been
obtained for the spin expectation value
$P(t)\stackrel{\rm def}{=}\langle S_z(t) \rangle$.
At the Toulouse point
$\Jpar^{\rm TP}/2\pi v_F=1-1/\sqrt{2}$ \cite{Toulouse} $P(t)$
can be evaluated exactly
\cite{Leggett} and one finds a purely exponential decay
$P(t)=\frac{1}{2}\exp(-2t/\pi w t_{\rm K})$. Here and in the sequel,
we define the Kondo time scale as $t_{\rm K}=1/T_{\rm K}$ with
the Kondo temperature defined via the impurity contribution to
the Sommerfeld coefficient $\gamma_{\rm imp}=w\pi^2/3T_{\rm K}$,
where $w=0.4128$ is the Wilson number. Using the form factor
approach Lesage and Saleur could derive exact results for $P(t)$ even
away from the Toulouse point \cite{Lesage}. For $0<\Jpar<\Jpar^{\rm TP}$
they find that the spin expectation value again decays exponentially
for large times $t/t_{\rm K}\gg 1$ with the same exponential dependence
$P(t)\propto \exp(-2t/\pi w t_{\rm K})$, however,
its behavior at finite times is more complicated.

Since $P(t)=2C(t_w=0,t)$ these results raise the question of
how the crossover between non--equilibrium exponential decay
and equilibrium algebraic decay occurs as a function of the
waiting time~$t_w$.

\textit{Toulouse point.--}
We have addressed this issue by evaluating
$C_{\rm I,II}(t_w,t)$ exactly at the Toulouse point
of the model. One finds that the result is the same in both situation
I) and II):
\bea
C_{\rm I,II}(t_w,t)&=&C_{\rm eq}(t) \nn \\
&&-2e^{-t_w/t_{\rm B}} s(t)\big(
s(t_w) e^{-t/t_{\rm B}}-s(t_w+t)\big)  \nn \\
&&-e^{-2t_w/t_{\rm B}} \big(
s(t_w) e^{-t/t_{\rm B}}-s(t_w+t)\big)^2
\label{C_nonequ}
\eea
for $t,t_w\geq 0$.
Here $s(\tau)=(t_{\rm B}/\pi) \int_0^\infty d\omega\,
\sin(\omega\tau)/(1+\omega^2 t_{\rm B}^2)$
with the abbreviation $t_{\rm B}\stackrel{\rm def}{=}\pi w t_{\rm K}$.
In terms of $s(\tau)$ the equilibrium correlation function
reads
\beq
C_{\rm eq}(t)=\frac{1}{4}\,e^{-2t/t_{\rm B}}-s^2(t) \ .
\label{C_eq}
\eeq
Notice that $s(t)\simeq t_{\rm B}/\pi t$ for $t\gg t_{\rm B}$ leading to
the algebraic long-time decay. Therefore
the Fourier transform of the equilibrium correlation function
is proportional to $|\omega|$ for small
frequencies $\omega/T_{\rm K}\ll 1$:
$C_{\rm eq}(\omega)\propto |\omega|/T_{\rm K}^2$.

\begin{figure}[tb]
\begin{center}
\includegraphics[width=0.45\textwidth,clip]{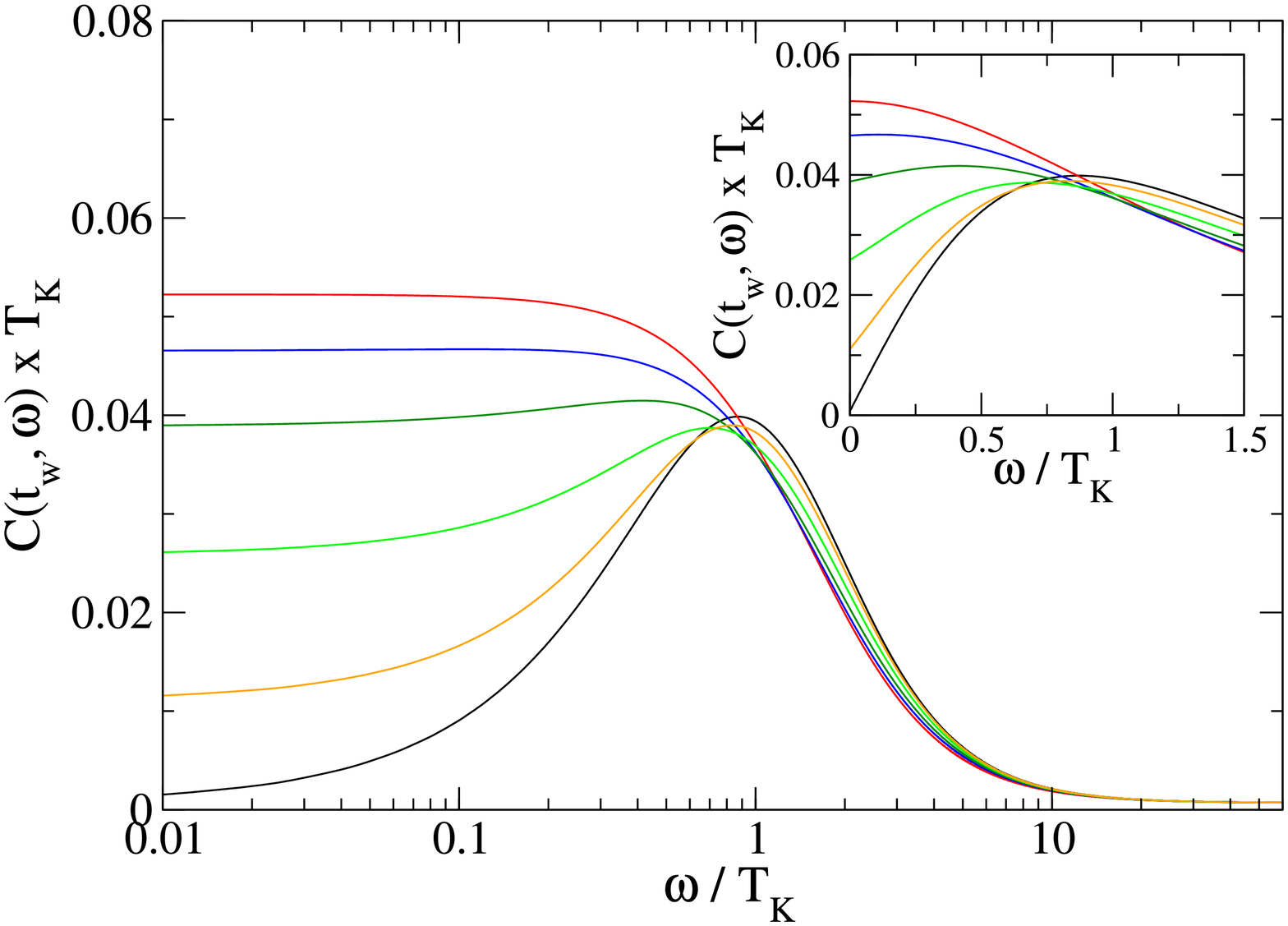}
\end{center}
\vspace*{-0.6cm}
\caption[]{Universal curves for the spin--spin correlation function
$C(t_w,\omega)$ at the Toulouse point for various waiting times
($t_w=0, t_{\rm K}/4, t_{\rm K}/2, t_{\rm K}, 2t_{\rm K}, \infty$ from top to bottom).
The inset depicts the same curves on a linear scale, which shows
more clearly
the onset of the non--analyticity ($C(t_w,\omega)-C(t_w,0)\propto |\omega|$)
for waiting times $t_w>0$ leading to the algebraic long--time decay.
\label{fig_TP}}
\end{figure}

For zero waiting time $t_w=0$ the non-equilibrium correlation function
Eq.~(\ref{C_nonequ}) shows the
well--known purely exponential decay
$C_{\rm I,II}(t_w=0,t)=\frac{1}{4} e^{-2t/t_{\rm B}}$,
while for any fixed $t_w>0$ the algebraic long time behavior dominates with
an amplitude that is suppressed depending on the waiting time:
\beq
C_{\rm I,II}(t_w,t)\simeq -s^2(t)\,\left(1-e^{-t_w/t_{\rm B}}\right)^2
+O\left(\left(t_{\rm B}/t\right)^3\right)
\label{CI_crossover}
\eeq
for $t\gg{\rm max}(t_w,t_{\rm B})$.
In particular, one can read off from (\ref{C_nonequ}) that
the difference between the non--equilibrium and equilibrium
correlation function decays exponentially fast as a function of
the waiting time
\beq
|C_{\rm I,II}(t_w,t) - C_{\rm eq}(t)| \propto e^{-t_w/t_{\rm B}}
\label{equilibration}
\eeq
for $t_w\gg t_{\rm B}$.
These results can be understood by noticing that the initial
state corresponds to an excited state of the model,
therefore yielding a different spin dynamics from equilibrium.
After a time scale of order~$t_{\rm K}$ corresponding to the
low--energy scale of the model, the initial non--equilibrium state
leads to the behavior of the equilibrium ground state with
deviations that decay exponentially fast as one increases the
waiting time \cite{note}.
This crossover behavior of the non-equilibrium correlation
function at the Toulouse point is shown in Fig.~\ref{fig_TP}
for the one--sided Fourier transform
with respect to the time difference~$t$
\beq
C(t_w,\omega)=\int_{-\infty}^\infty \frac{dt}{2\pi}\: e^{i\omega t}\:
C(t_w,|t|) \ .
\eeq

\textit{Method.--}
We perform the standard procedure of bosonizing the Kondo Hamiltonian
in terms of spin--density excitations $b^\pdag_k, b^\dag_k$,
and then eliminating its $\Jpar$--coupling using a polaron transformation
$
U=\exp[i \delta \,S_z\,\sum_{k>0} (1/\sqrt{k})\,
(b^\pdag_k-b^\dag_k)]
$
(for details see e.g.\ \cite{Leggett}).
At the Toulouse point the transformed Hamiltonian $\tilde H=U^\dag H U$
can be refermionized \cite{Leggett}
\beq
\tilde H=\sum_k \epsilon_k \Psi^\dag_k \Psi^\pdag_k + \sum_k V_k
\left(\Psi^\dag_k d + d^\dagger \Psi^\pdag_k \right) \ .
\label{Hamiltonian_RLM}
\eeq
Here the spinless fermions $\Psi(x)$ correspond to
soliton excitations built from the bosonic spin--density waves
with the refermionization identity
$\Psi(x)\propto\exp\big[i\sum_{k>0}(1/\sqrt{k})\,
(b^\pdag_k e^{-ikx}-b^\dag_k e^{ikx})\big]$.
Eq.~(\ref{Hamiltonian_RLM}) can be interpreted as a resonant
level model with the hybridization function
$\Delta(\epsilon)\stackrel{\rm def}{=}\pi\sum_k V_k^2 \delta(\epsilon-\epsilon_k)
=T_K/\pi w$
and a fermionic impurity orbital $d$ with the identity
$S_z=d^\dagger d-1/2$.
Model~I is therefore represented by a resonant level
model with a time--dependent impurity orbital energy
$\epsilon_d(t)=h(t)$
\bea
\tilde H_{\rm I}(t)&=&\sum_k \epsilon_k \Psi^\dag_k \Psi^\pdag_k
+ \epsilon_d(t)\,(d^\dag d-1/2) \nn \\
&&+ \sum_k V_k \left(\Psi^\dag_k d + d^\dag \Psi^\pdag_k \right)
\eea
with $\epsilon_d(t<0)=\infty$, $\epsilon_d(t\geq 0)=0$.
In model~II the spin is not coupled to the Fermi sea for time $t<0$: 
this leads to a time--dependent hybridization function
and a time--dependent potential scattering term (due to the 
polaron transformation $U$)
\bea
\tilde H_{\rm II}(t)&=&\sum_k \epsilon_k \Psi^\dag_k \Psi^\pdag_k
+ \sum_k V_k(t) \left(\Psi^\dag_k d + d^\dag \Psi^\pdag_k \right)
  \nn \\
&&+\langle S_z(t=0)\rangle \sum_{k,k`} g_{kk'}(t)
(\Psi^\dag_k \Psi^\pdag_{k'}-1/2)
\label{HII}
\eea
with $g_{kk'}(t<0)=(1-\sqrt{2})/2\pi v_F$,
$g_{kk'}(t\geq 0)=0$ and $\Delta(\epsilon;t<0)=0$,
$\Delta(\epsilon;t\geq 0)=T_K/\pi w$.

In order to evaluate the non--equilibrium spin dynamics we 
use the quadratic form of $\tilde H_{\rm I,II}$ to solve 
the Heisenberg equations
of motion for the operators $d^\dagger(t), d(t)$ exactly.
After some straightforward
algebra one can write the operator $S_z(t)$ for $t>0$
as a quadratic expression in
terms of the operators $\Psi^\dag_k(t=0)$,
$\Psi^\pdag_k(t=0)$, $d^\dag(t=0)$ and $d(t=0)$ at time $t=0$
\cite{long_paper}. The non--equilibrium correlation function
is then given by
\beq
C_{\rm I,II}(t_w,t)= \frac{1}{2}
\langle \tilde G_{\rm I,II}| \{S_z(t_w),S_z(t_w+t)\}|\tilde G_{\rm I,II}
\rangle \ .
\label{Heisenberg_C}
\eeq
where we insert the solution of the Heisenberg equation of
motion for $S_z(t)$.
The initial non--equilibrium states $|\rm \tilde G_{\rm I,II}\rangle$
remain time--independent in the Heisenberg picture and
are simply the ground states of the Hamiltonians
$\tilde H_{\rm I,II}(t)$ for time $t<0$.  
Inserting these expressions in (\ref{Heisenberg_C}) yields
(\ref{C_nonequ}) after some tedious but straightforward algebra \cite{long_paper}.
For completeness we also
give the result for the imaginary part of the non--equilibrium
Greens function
\begin{eqnarray}
&&\!\!\!
K_{\rm I,II}(t_w,t) \stackrel{\rm def}{=} \frac{1}{2}\,
\langle\, [S_z(t_w),S_z(t_w+t)]\, \rangle_{\rm I,II}  \\
&&\!\!\! =-i\,e^{-t/t_{\rm B}} (s(t)-s(t_w+t)e^{-t_w/t_{\rm B}}
+s(t_w)e^{-(t_w+t)/t_{\rm B}} ) \nn
\end{eqnarray}
which again approaches the equilibrium result
$K_{\rm eq}(t)=-i\,e^{-t/t_{\rm B}} s(t)$ exponentially fast
as a function of $t_w/t_{\rm B}$.

\textit{Kondo limit.--}
The Toulouse point exhibits many universal features of the
strong--coupling phase of the Kondo model like local Fermi
liquid properties, however,
other universal properties like the Wilson ratio 
depend explicitly on the coupling~$J_\parallel$.
This raises the question which of the above non--equilibrium to
equilibrium crossover properties are generic
in the strong--coupling phase. We investigate this question
by using the flow equation method \cite{Wegner} that allows us to
extend our analysis away from the Toulouse point in a controlled
expansion. In this Letter we focus on the 
experimentally most relevant
limit of small Kondo couplings (notice that the flow equation approach
is not restricted to this limit and can
be used for general $\Jpar$ \cite{prb}).

The flow equation method diagonalizes a many--particle
Hamiltonian through a sequence of infinitesimal unitary
transformations in a systematic approximation \cite{Wegner}. 
This approach was carried through
for the Kondo model in Ref.~\cite{Hofstetter}. Since the
Hamiltonian is transformed into its diagonal basis, we 
can follow the same steps 
as in the Toulouse point analysis: i)~The Heisenberg equations
of motion for the unitarily transformed observables can
be solved easily with respect to the diagonal Hamiltonian. 
ii)~One then re-expresses the time--evolved
operators through the operators in the initial (non--diagonal)
basis for time $t=0$. 
iii)~The correlation functions (\ref{Heisenberg_C}) for
general $t_w$ are evaluated.
An operator product expansion to leading order is employed like
in Ref.~\cite{Hofstetter} to close the resulting systems of
equations. 

The above procedure can be used quite generally to apply the flow
equation method to time--dependent Hamiltonians of the above kind. For the
Kondo model specifically, one can simplify the calculation by
using the results from Ref.~\cite{Slezak}: it was shown 
that a resonant level
model (\ref{Hamiltonian_RLM}) with a universal non--trivial hybridization
function $\Delta_{\rm eff}(\epsilon)\neq$~const.\ can be used as an
effective model for the spin dynamics on all time scales;
the only free parameter is the low--energy scale~$T_K$.
Similarly, the effective Hamiltonian (\ref{HII}) 
with $\Delta(\epsilon;t>0)=\Delta_{\rm eff}(\epsilon)$
from Ref.~\cite{Slezak} yields 
the $S_z$--spin dynamics for both non--equilibrium situations I and~II.
A careful analysis \cite{prb} shows that the only
effect not captured by the 
resonant level model is the polaron--like
transformation that is contained in the complete flow
equation approach. This polaron tranformation leads to an
initial potential scattering term like in (\ref{HII}) with
$g_{kk'}(t<0)=(\lambda(B_{\rm eff})-\sqrt{2})/2\pi v_F$,
where $B_{\rm eff}=1/(\epsilon_k^2+\epsilon_{k'}^2)$ and
$\lambda(B)$ is the flowing scaling dimension according to
Ref.~\cite{Hofstetter} ($\lambda(B=0)=\sqrt{2}$ and
$\lambda(B\rightarrow\infty)=1$). However, this
initial potential scattering term has a neglible effect (relative
error $<5\%$) unless one is interested in 
short waiting times $t_w\lesssim t_{\rm K}/4$: for larger waiting times
the effective Hamiltonian from Ref.~\cite{Slezak} yields a very
accurate description without the need for the full flow equation
analysis.

\begin{figure}[tb]
\begin{center}
\includegraphics[width=0.45\textwidth,clip]{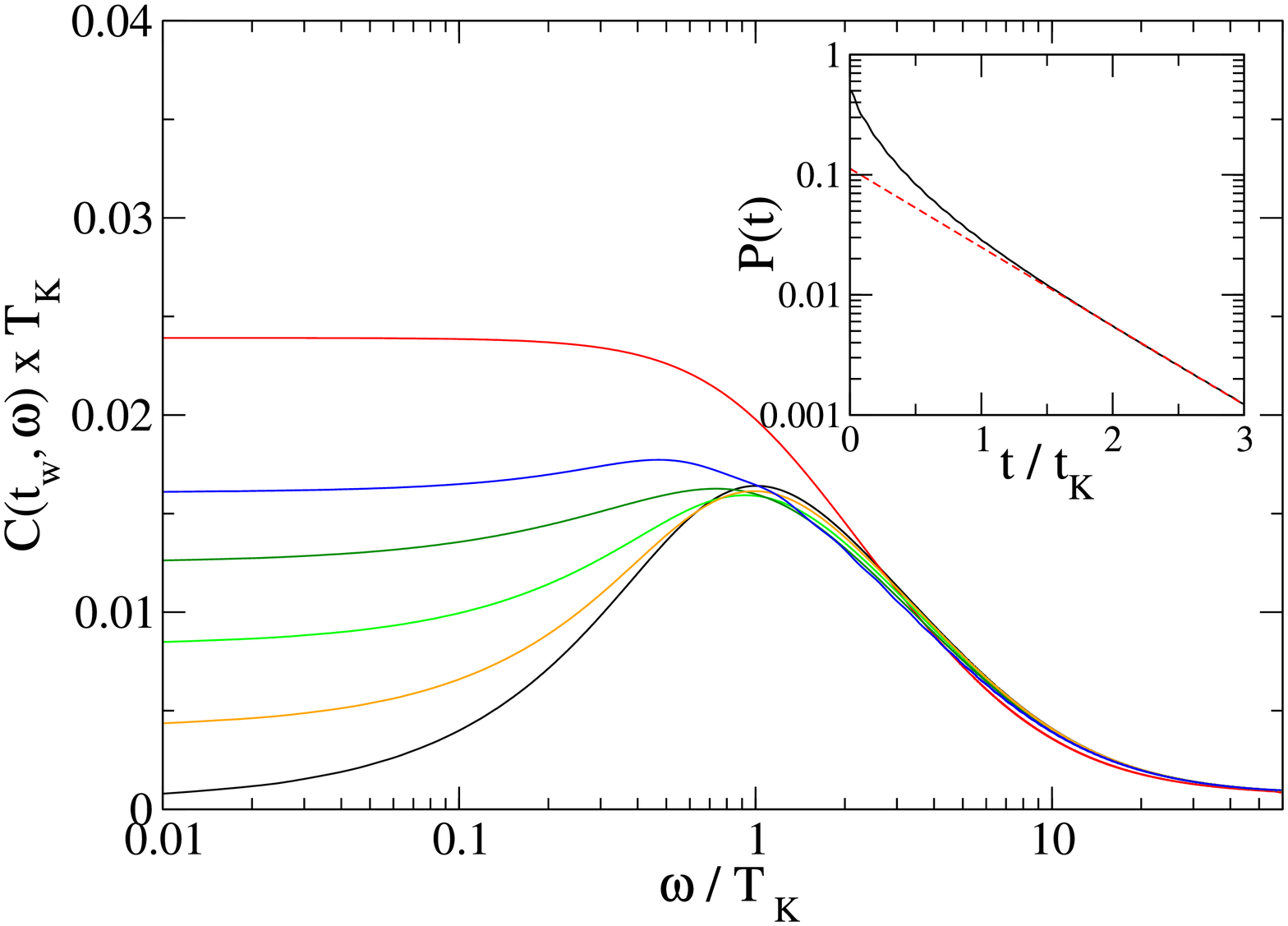}
\end{center}
\vspace*{-0.6cm}
\caption[]{Universal curves for the spin--spin correlation function
$C(t_w,\omega)$ in the limit of small Kondo couplings (Kondo limit)
for various waiting times
($t_w=0, t_{\rm K}/4, t_{\rm K}/2, t_{\rm K}, 2t_{\rm K}, \infty$ from top to bottom
like in Fig.~\ref{fig_TP}).
The inset shows the spin expectation value $P(t)$: the dashed
line is an asymptotic fit $P_{\rm asym}(t)=0.11\exp(-1.51\,t/t_{\rm K})$.
\label{fig_KL}}
\end{figure}

From the quadratic Hamiltonian (\ref{HII}) with the above effective
coupling constants (including the initial potential scattering term)
one can easily evaluate the non--equilibrium
correlation functions; results are depicted in Fig.~\ref{fig_KL}.
The key observations from the Toulouse point analysis hold
in the Kondo limit as well, only the crossover behavior is more
complicated: 1)~The system approaches equilibrium behavior
exponentially fast as a function of $t_w/t_{\rm K}$ (compare
Eq.~(\ref{equilibration})).
Notice that the initial approach for small $t_w$ in Fig.~\ref{fig_KL}
is faster than
at the Toulouse point (Fig.~\ref{fig_TP}).
2)~An algebraic long--time decay~$\propto t^{-2}$ dominates
for all nonzero waiting times $t_w>0$ (compare Eq.~(\ref{CI_crossover})).

For zero waiting time $t_w=0$ the inset in Fig.~\ref{fig_KL}
shows the decay of the spin expectation value $P(t)$ in the
Kondo limit, which has not been previously calculated explicitly on all time
scales. For large~$t/t_{\rm K}$ the behavior crosses over into an exponential
decay which agrees very well with the exact asymptotic
result from Ref.~\cite{Lesage}: 
$P_{\rm exact}(t)\propto\exp(-2t/\pi w t_{\rm K})\propto\exp(-1.54t/t_{\rm K})$. 
On shorter
time scales the decay is faster, which is due to unrenormalized
coupling constants at large energies that dominate the short--time
behavior. 

\textit{Conclusions.--}
Summing up, we have investigated the crossover to
equilibrium behavior for a Kondo model that is prepared in
an initial non--equilibrium state. We calculated
the non--equilibrium spin--spin correlation function on all time scales
and could show that it evolves exponentially fast
towards its equilibrium form for large waiting time of the first
spin measurement $t_w\gg t_{\rm K}$, see Eq.~(\ref{equilibration}).
Our results also established that the flow equation method is
a very suitable approach for studying such non--equilibrium problems:
it agrees with very good accuracy with exact results for
both $t_w=0$ and $t_w=\infty$, and can describe the crossover
regime as well.

Finally, it is worthwile to
recall the fundamental quantum mechanical observation that the
overlap between the time--evolved non--equilibrium
state and the true ground state of the Kondo model is always
{\em time--independent}. Therefore it is not strictly accurate to conclude
from our results that an initial non--equilibrium state "decays"
into the ground state: rather,
quantum observables which exhibit
equilibration behavior are probes for which the time--evolved
initial non--equilibrium state eventually "looks like" the ground state.
Since the notion of equilibration into the ground state
plays a fundamental role in quantum physics, it would be very interesting to study
other systems with quantum dissipation to see which
of the crossover and equilibration properties derived in this Letter are generic.

We acknowledge valuable discussions with T.~Costi and D.~Vollhardt.
This work was supported
through SFB~484 of the Deutsche Forschungsgemeinschaft (DFG).
SK acknowledges support through a Heisenberg fellowship 
of the DFG.


\begin{thebibliography}{11}
\bibitem{note}
Notice that it is not possible to think of
the time--evolved initial non--equilibrium state as effectively
being equivalent to an equilibrium system
with a nonzero temperature that depends on the
waiting time $t_w$: nonzero temperature leads to an exponential
decay  and not to the correct algebraic
long--time behavior for $t_w>0$ in (\ref{CI_crossover}).
\bibitem{prb}
Details of the flow equation analysis will be published separately.
\end{thebibliography}
\end{document}